\documentclass[reprint,amsmath,amssymb,aps]{revtex4-1}
\pdfoutput=1

\usepackage{graphicx}
\usepackage{dcolumn}
\usepackage{bm}

\begin{document}

\preprint{APS/123-QED}

\title{A micro dalton resolution mass sensor using optically cooled
microdisks via short-time measurement}
\thanks{}%

\author{Jian Liu}
 \altaffiliation[]{}
\author{Ka-Di Zhu}%
 \email{zhukadi@sjtu.edu.cn}
\affiliation{Key Laboratory of Artificial Structures and Quantum Control (Ministry of
Education), School of Physics and Astronomy, Shanghai Jiao Tong
University, 800 DongChuan Road, Shanghai 200240, China,
Collaborative Innovation Center of Advanced Microstructures, Nanjing, China
}%

\date{\today}

\begin{abstract}
We propose a mass sensor using optically trapped and cooled dielectric
microdisks with "measuring after cooling" scheme. The center-of-mass motion
of a trapped particle in vacuum can experience extremely low dissipation
resulting in robust decoupling from the heat bath. Thus the measurement can
be accomplished in a short time before a jump in the phonon number after the
optomechanical cooling is finished. This method can lead to $10^{-6}$ dalton
mass sensitivity which is a thousand times smaller than the mass of one
electron.

\end{abstract}

\pacs{Valid PACS appear here}
\maketitle


\section{Introduction}

Mechanical resonators are widely used as inertial balances to detect small
quantities of adsorbed mass due to the shifts in their vibrational
frequencies. Nanoelectromechanical systems have been proposed for highly
sensitive mass detection of neutral species. Significant progresses have
been made by using nanofabricated resonators[1,2] and carbon nanotubes[3-5]
for mass sensing. However, so far, the best mass resolution achieved with
microfabricated resonators has been 200yg in the base pressure of $10^{-10}$%
Torr[3,4], whereas carbon nanotube resonators have achieved mass resolutions
as low as 2yg in base pressure of \ $3\times 10^{-11}$mbar[5]. Here we
demonstrate an optically levitated microdisk based optomechanical cooling
with $10^{-6}$Da resolution. The masses of neutral atoms, electron, neutrons
and other chemical molecules and biological molecules can be measured in
all-optical domain, which have applications in mass spectrometry[6],
magnetometry[7] and surface science[8].

Quantum optomechanical techniques in both the optical and microwave regimes
will provide motion and force detection near the fundament limit imposed by
quantum mechanics[9,10]. Optomechanical quantum control requires the
mechanical oscillator to be in or near its quantum ground state. Therefore,
nowadays the investigations of optomechanical cooling become a research
focus. The two main methods being pursued in optomechanics are
\textquotedblleft backaction cooling\textquotedblright\ which requires the
stable operation of a high-finesse cavity[11-14], and \textquotedblleft
feedback cooling\textquotedblright\ which requires precision measurement and
feedback[15-17]. Recently much effort has been directed toward optically
levitating nano- and micro-mechanical oscillators in ultrahigh vacuum such
as nanospheres[14,17], nanodiamonds[18,19], microdisk[20,21], and even the
living organisms[22]. The laser trapped objects has no physical contact to
the environment, leads to ultralow mechanical damping which makes it a
promising system for ground state cooling even at room
temperatures[14,17,20].

However, the optomechanical cooling will increase the mechanical damping
rate and cause the extra damping, thus will dramatically increase the
linewidth of the oscillator[23]. This effect can significantly decrease the
precision which strongly depends on the linewidth on spectrum. Here we
propose a unique approach toward this problem, wherein the detection can be
accomplished by a short integration time after the ground state cooling. An
optically trapped microdisk in vacuum is well isolated from the thermal
environment and can have a mechanical damping $\sim 10^{-7}Hz$. In this
case, the thermal decoherence rate reduces notably, thus the detection can
be done before the mechanical oscillator are heating out of the ground state
in the absence of cooling scheme. In addition, the decoherence and heating
rates are fundamentally limited by the momentum recoil of scattered photons
and can be reduced simply by using high finesse cavity and increasing the
trapping frequency[14,24]. Based on the "measuring after cooling" technique,
we theoretically show that the mass sensitivity can be improved by 6-7
orders than the traditional nanomechanical resonators[3-5], reaching $%
10^{-6} $Da resolution. Mass spectrometers actually measure mass-to-charge
ratios. Here the true mass of the atoms or molecules can, in principle, be
determined without any need to ionize, even the mass of the electrons can
also be determined. This unprecedented level of sensitivity allows us to
measure precisely the mass of neutral particle and to distinguish between
isotopes in inertial mass spectrometry measurements.

\section{Method}

Our approach involves optically trapping and cooling a cleanroom fabricated
microdisk without clamping to the substrate at room temperature. We choose a
silicon nitride microdisk of radius $r=1\mu m$ and thickness $b=50nm$. The
ultralight weight of the microdisk($\sim 10^{-16}kg$) may allow an ideal
situation in which a freely levitated disk is trapped against the gravity in
the anti-node of a strong optical standing wave. The light fields of
wavelength $\lambda =1\mu m$ is used for trapping. A second field is used to
cool and to measure the center-of-mass motion of the disk as shown in the
Fig.1. Following Ref.[17], we can cool the levitated microdisk
parametrically by a feedback loop.
\begin{figure}[tbph]
\includegraphics[width=7.5cm]{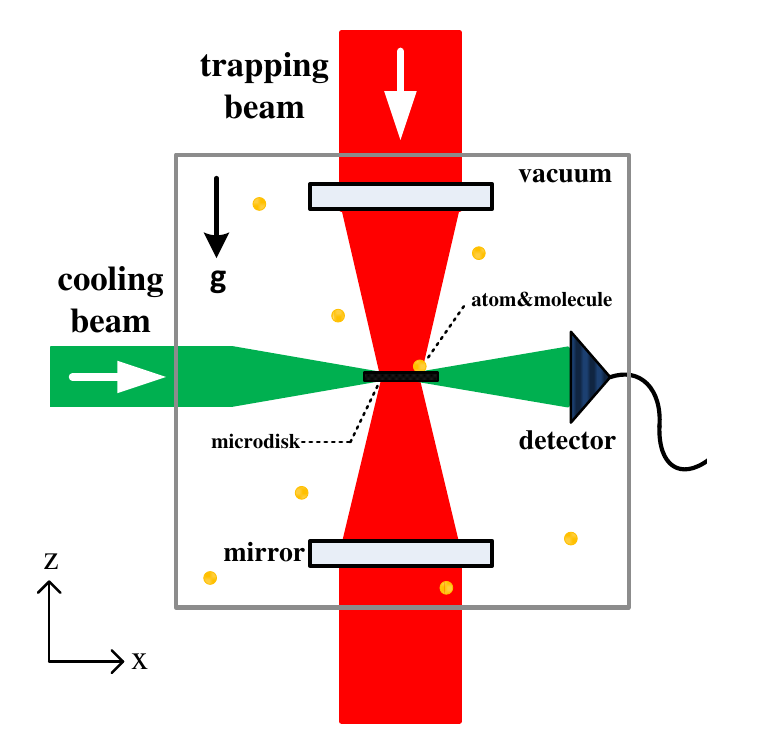}
\caption{Proposed experimental geometry. A silicon nitride microdisk is
trapped by the anti-node of a strong optical standing wave in the
Fabry-Perot cavity. The second light is used to cool and to measure the
center-of-mass motion of the levitated oscillator. The sensor can detect the
mass change induced by the external particles landing onto the surface of
the trapped microdisk.}
\end{figure}
Nanomechanical resonators can function as precision mass sensors because
their resonant frequency which is related to their mass will shift when a
particle accretes with the resonator. We expect that the sensor can detect
the mass change induced by the external particles landing onto the surface
of the trapped microdisk. The optically tapped microdisk can be described by
a harmonic oscillators with an effective mass $m$. For small oscillation
amplitudes, the nonlinearity is negligible and the three spatial dimensions
are decoupled. Specifically, we consider the vibration along x-axis which is
defined by the resonance frequency $\omega _{m}$. The frequency shift is
directly proportional to the additional or reduced mass.%
\begin{equation}
\delta m=-\frac{2m}{\omega _{m}}\delta \omega .
\end{equation}%
Mass sensing is based on monitoring the frequency shift $\delta \omega $ of $%
\omega _{m}$.

The scheme is separated into two step process, cooling and measurement.

1)cooling. The equation of motion for the microdisk's motion in $x$
direction (polarization direction) is
\begin{equation}
\overset{\cdot \cdot }{x}(t)+\gamma \overset{\cdot }{x}(t)+\omega
_{m}^{2}x(t)=\frac{1}{m}[\xi (t)+F_{opt}(t)].
\end{equation}%
Here $\xi (t)$ denoting fluctuating forces acting on the microdisk due to
the impact of the air molecules. For white noise, one has that $\left\langle
\xi (t)\xi (t^{\prime })\right\rangle =2k_{B}Tm\gamma \delta (t-t^{\prime })$
according to the fluctuation dissipation theorem, where $k_{B}$ is the
Boltzmann constant and $T$ is the temperature of the chamber. The
oscillator's damping rate can be written as $\gamma =\gamma _{m}+\gamma
_{fb} $, where $\gamma _{m}$ accounts for the interaction with the
background gas and $\gamma _{fb}$ is the damping introduced by feedback
cooling. $F_{opt}(t)=\Delta k_{cool}(t)x(t)$ is a time-varying,
nonconservative optical force introduced by parametric feedback, here $%
\Delta k_{cool}(t)$ is the additional trapping stiffness. Thus, feedback
cooling leads to shifts $\gamma _{fb}$ and $\Delta \omega $ in the disk's
natural damping rate $\gamma _{m}$ and oscillation frequency $\omega _{m}$.
Activation of the parametric feedback loop gives rise to the additional
damping, and predicts that the CM temperature is reduced down to%
\begin{equation}
T_{CM}=T\frac{\gamma _{m}}{\gamma _{m}+\gamma _{fb}},
\end{equation}

2)measurement. To accurately measure the microdisk's frequency, we need to
decrease the CM temperature. This can only be accomplished by using the
cooling laser and hence an additional optical feedback damping. On the other
hand, high feedback damping gives rise to large oscillation linewidth and
significantly decreases the sensitivity. Thus we will not apply the
parametric feedback in the measurement. One can stop feedback loop by
ceasing the modulation in the cooling laser intensity at a time $t^{\prime }$%
, thus we have%
\begin{equation}
F_{opt}(t)=0\text{ \ \ for }t>t^{\prime }.
\end{equation}%
It leads to shifts $\gamma _{fb}=0$, and the total damping reduces to $%
\gamma _{m}$. Thus we only analyze here the heat of the mechanical motion of
the center-of-mass of a dielectric disk due to the gas pressure and the
photon recoil.

The variance of the position can be computed by solving the differential
stochastic equation. We suppose $F_{opt}(t)=0$ in the Eq.(2) and consider $%
\omega _{m}\gg \gamma _{m}$ which is always fulfilled in the levitating
resonators, then we have%
\begin{equation}
\left\langle \lbrack x(t)-\left\langle x(t)\right\rangle ]^{2}\right\rangle
\approx \frac{k_{B}T}{m\omega _{m}^{2}}[1-e^{-\gamma _{m}t}].
\end{equation}%
By considering the equipartition principle, the variance allows us to
compute the increase of energy by taking $\Delta E(t)=m\omega
_{m}^{2}\left\langle [x(t)-\left\langle x(t)\right\rangle ]^{2}\right\rangle
$. Hence one can compute the time $\Delta t$ required to increase one
quantum $\hbar \omega _{m}$ of energy in the quantum harmonic oscillator. In
the case of $\hbar \omega _{m}\ll k_{B}T$, the time $\Delta t$ is given by
solving $\Delta E(\Delta t)=\hbar \omega _{m}$ and reads%
\begin{equation}
\Delta t=\frac{1}{\gamma _{m}}\log (1-\frac{\hbar \omega _{m}}{k_{B}T}%
)\approx \frac{\hbar \omega _{m}}{k_{B}T\gamma _{m}}=\frac{1}{\Gamma _{th}}.
\end{equation}%
Here $\Gamma _{th}=k_{B}T\gamma _{m}/\hbar \omega _{m}$ is often referred to
as the thermal decoherence rate and given by the inverse time it takes for
one quantum to enter from the environment. The recoil imparted on the
particle by a scattered photon is a small effect since the photon momentum
is small compared to the momentum of the mechanical oscillator. However, at
very low temperatures and pressures, collisions with air molecules become
negligible and photon recoil takes over as the dominating decoherence
process. Considering that the strong scattering introduces recoil heating,
one can obtain the total decoherence rate $\Gamma ^{\prime }=\Gamma
_{th}+\Gamma _{recoil}$, and the total heating time%
\begin{equation}
\Delta t^{\prime }=\frac{1}{\Gamma ^{\prime }}.
\end{equation}%
Hence, we can choose a measurement time $\tau $ that is sufficiently short
to ensure that the measurement can be accomplished before a single phonon is
exchanged with the thermal bath in the absence of cooling scheme.

We now turn to the evaluation of the minimum measurable frequency shift, $%
\delta \omega $, limited by thermomechanical fluctuations of a levitated
resonator readout by a probe laser. In such a measurement, the resonator is
driven at a constant mean square amplitude, $x_{c}$, by the harmonic
trapping potential. An estimate for $\delta \omega $ can be obtained by
integrating the weighted effective spectral density of the frequency
fluctuations by the normalized transfer function of the measurement loop.
Performing this integration for the case where $Q\gg 1$, we obtain[25]
\begin{equation}
\delta \omega =\sqrt{\frac{\gamma _{m}k_{B}T_{CM}\Delta f}{k_{trap}x_{c}^{2}}%
}.
\end{equation}%
Here, $\Delta f$ is the measurement bandwidth. Experimentally, the
measurement time $\tau $ is defined by the detection bandwidth $\tau =1/2\pi
\Delta f$.

If the detection can be completed in a short time, fulfilling that $\tau
\leqslant \Delta t^{\prime }$, the phonon number can remain unchanged after
the feedback cooling is stopped. It allows the measurement in ground state
and the ultralow CM temperature $T_{CM}$ can be achieved. To obtain the
minimum detectable mass, we rewrite Eq.(8) using the expression for $\delta
m $ given in Eq.(1) supposing $\tau =\Delta t^{\prime }$.%
\begin{equation}
\delta m=\frac{1}{\omega _{m}^{2}}\sqrt{\frac{2m\gamma _{m}k_{B}T_{CM}}{\pi
x_{c}^{2}\Delta t^{\prime }}}.
\end{equation}

\section{Sensitivity}

In equilibrium, the microdisk is situated at $z=0$, in the anti-node of an
optical standing wave $E(z)\propto \cos kz$ (see Fig.1). The optical field
polarizes the dielectric disk, yielding a gradient force trap around the
equilibrium position. In the absence of any internal forces, the optical
field traps a thin disk ($d\ll \lambda $) with a restoring frequency given by%
\begin{equation}
\omega _{m}=\left[ \frac{2k^{2}I_{t}(\varepsilon _{r}-1)}{\rho c}\right]
^{1/2}.
\end{equation}%
Here $I_{t}$ is trapping beam intensity, $k=2\pi /\lambda $ is the optical
wavevector, $\rho $ is the mass density and $\varepsilon _{r}$ is the
dielectric constant. For the silicon nitride microdisks with the dielectric
constant $\varepsilon _{r}=4$, $\rho =2700kg/m^{3}$ we obtain the trapping
frequency along x-axis $\omega _{m}=54MHz$ with $I_{t}=10W/\mu m^{2}$. The $%
x_{c}$ is the maximum root mean square(rms) amplitude produced a
predominantly linear response. For a Gaussian field distribution, the
nonlinear coefficients are given by $\xi =-2/W^{2}$[26], Considering the
beam waist radius $W=r=1\mu m$, for small displacements, $\left\vert
x_{c}\right\vert \ll \left\vert \xi \right\vert ^{-1/2}=0.7\mu m$, the
nonlinearity is negligible. In our considerations, $x_{c}$ is taken to be
about one orders of magnitude smaller, we choose $x_{c}=50nm$. Consider the
mass of the microdisk $m=4.2\times 10^{-16}kg$, one can obtain the trapping
stiffness $k_{trap}=m\omega _{m}^{2}=1.22N/m$.

A laser-trapped microdisk in ultrahigh vacuum, by contrast, has no physical
contact to the environment. Thus, the mechanical damping of a levitated
microdisk $\gamma _{m}$ is limited only by collisions with residual air
molecules. From kinetic theory we find that[21]%
\begin{equation}
\gamma _{m}=\frac{32P}{\pi v\rho b},
\end{equation}%
where $P$ is the pressure, $v=\sqrt{3k_{B}T/m_{gas}}$ and $m_{gas}$ are the
root-mean-square velocity and mass of gas molecules, respectively. In order
to achieve ultralow damping rate, the measurement must be completed in
ultrahigh vacuum. The measurement set-up in Ref.[5] was prepared by lowering
the base pressure to $3\times 10^{-11}$mbar to minimize adsorption of
unwanted molecules. Let us choose the almost same parameter in simulation,
supposing $P=10^{-11}$mbar, then we obtain $\gamma _{m}=1.5\times 10^{-7}Hz$%
and the thermal decoherence rate $\Gamma _{th}=0.12Hz$ with room temperature.

We have assumed that the trapping beam has a Gaussian profile. For cavity
optomechanics, it will be necessary to trap the microdisk within a
Fabry--Perot cavity, as illustrated in Fig.1. We now consider the effect of
photon recoil heating. The number of coherent oscillations before a jump in
the phonon number can be written as[20]%
\begin{equation}
N=\frac{1}{2\pi }\frac{V}{V_{c}}\frac{\omega _{c}}{k}\frac{\omega
_{m}^{2}\rho c}{k^{2}I_{t}},
\end{equation}

Here, $\omega _{c}=ck$, $V$ is the volume of the disk and $V_{c}=\pi
W^{2}L/4 $ is the cavity mode volume. We assume a high finesse cylinder
cavity of length $L=1cm$, and finesse $F_{c}=3\times 10^{5}$[14], leading to
a cavity decay rate $\kappa =c\pi /2F_{c}L=1.6\times 10^{5}Hz$. Thus we
obtain $N=2.3\times 10^{5}$ for a trapped SiN microdisk with frequency $%
\omega =54MHz $. Then the photon recoil heating can be obtained by $\Gamma
_{recoil}=\omega _{m}/2\pi N=37.6Hz$. Plugging these values into the Eq.(7),
then we get the total heating time $\Delta t^{\prime }=26.5ms$.

According to Eq.(11), a low pressure implies a low damping rate and thus, we
find that lower CM temperature can be achieved at higher vacuum. Recalling
that the typical cooling rate is of $\gamma _{fb}\approx 20Hz$ in the
Ref.[17], one can cool its motional degrees of freedom from room temperature
to $T_{CM}=2.3\mu K$ with $P=10^{-11}$mbar according to Eq.(3). Note that,
in the photon dominated regime and in the absence of feedback cooling, the
depth of the trapping potential can be estimated by $\Delta U\approx
(1/2)k_{trap}x_{c}^{2}\sim 10^{-15}J$. The phonon energy $k_{B}T_{eff}\sim
10^{-29}J$, this energy is much smaller than the depth of the trapping
potential in our scheme, and therefore the particle is unlikely to escape as
it heats up without feedback control.
\begin{table}[tbp]
\begin{ruledtabular}
\caption{Parameters of the cooled levitated microdisk optomechanical
systems}
\begin{tabular}{lllll}
Parameter ~ & Units ~ & Value ~ \\ \hline
Radius of microdisk, $r$ & $\mu m$ & $1$\\
Thickness of microdisk, $b$ & $nm$ & $50$\\
Beam waist radius, $W$ & $\mu m$ & $1$\\
Trapping beam intensity, $I_{t}$ & $W/\mu m^{2}$ & $10$\\
Trapping beam wavelength, $\lambda$ & $\mu m$ & $1$\\
Air pressure, $P$ & $mbar$ & $10^{-11}$\\
Chamber temperature, $T$ & $K$ & $300$\\
Damping of feedback, $\gamma _{fb}$ & $Hz$ & $20$\\
Cavity finesse, $F_{c}$ & $null$ & $3\times 10^{5}$\\
Cavity length, $L$ & $cm$ & $1$\\
\end{tabular}
\end{ruledtabular}
\end{table}
We list all the fundamental parameters in the Table I. Following Eq.(9), one
can obtain the sensitivity of the mass probing $\delta m=1.5\times 10^{-33}kg
$, this value corresponds to $0.9\times 10^{-6}$Da and $0.84$keV in the
natural unit. We expect the ultrahigh resolution mass sensor can be achieved
under these conditions.

The major determinants of the mass sensitivity is the gas pressure in the
chamber. The lower pressure limit of sputter-ion pumps is in the range of $%
10^{-11}$mbar. Lower pressures in the range of $10^{-12}$mbar can only be
achieved when the sputter-ion pump works in a combination with other pumping
methods[27,28]. To investigate the impact of pressure, we depict the
sensitivity as a function of the pressure in Fig.2 by the red line.
According to the figure, higher mass sensitivity requires a lower pressure
and we find that micro dalton resolution can be achieved at ultrahigh vacuum
($10^{-11}$mbar).
\begin{figure}[tbp]
\includegraphics[width=8cm]{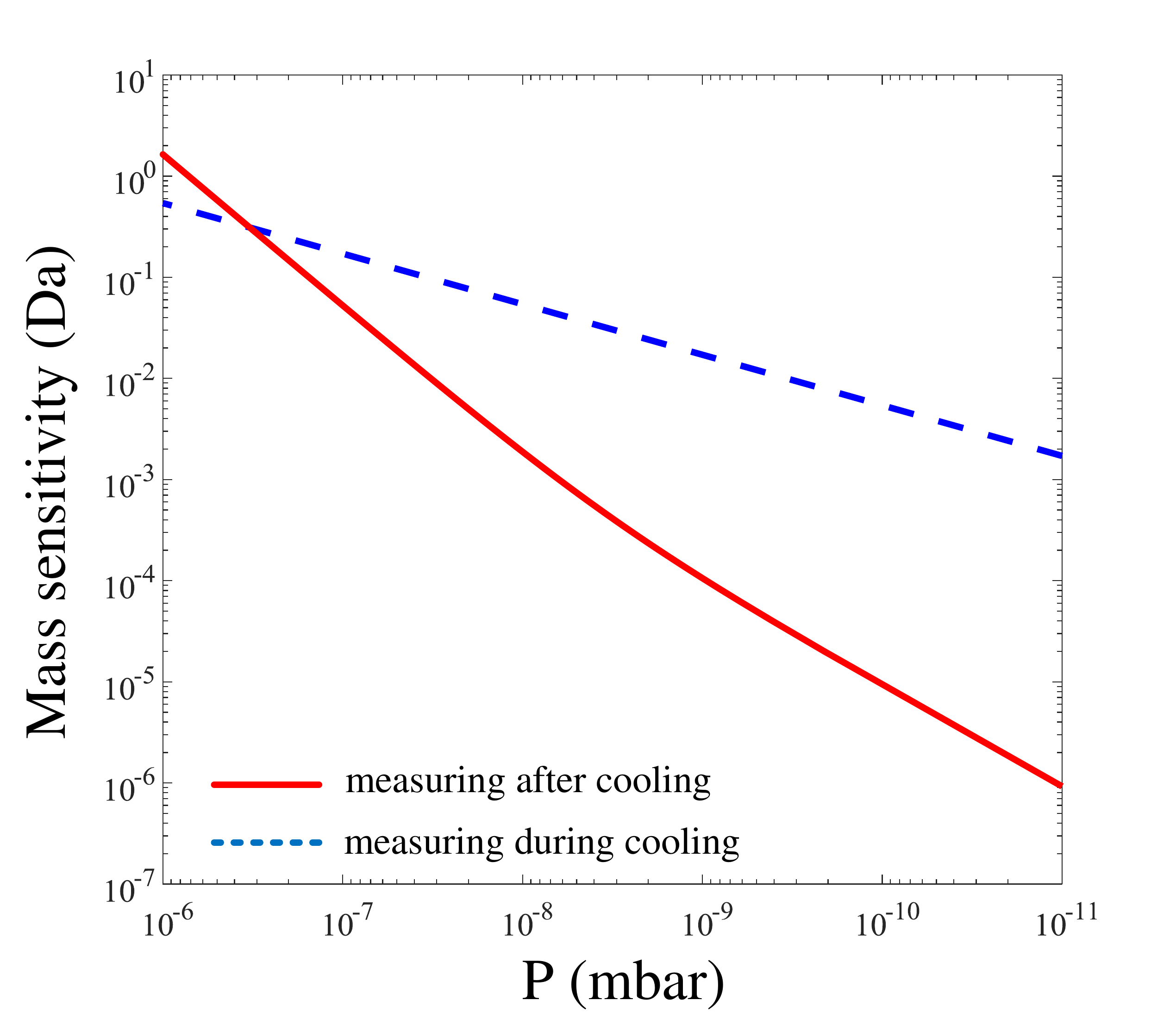}
\caption{Mass sensitivity as a function of pressure. The blue dash line
shows the result for the traditional "measuring in cooling" scheme with the
sample time of 1s and the red line shows the result of the "measuring after
cooling" scheme which can be completed in the heating time. The turning in
the red line is caused by the photon recoil noise. The parameters are
corresponding to Table I}
\end{figure}
Compared with the previous sensitive measurement with cooling[10], the
distinct difference of our scheme is the "measuring after cooling" method.
We first radiate a cooling field on the optically levitated system. Once the
trapping microdisk is cooled to the ground state, then we stop cooling and
begin to measure, the masses can be detected with a short integration time
before a jump in the phonon number, thus the CM temperature keeps the same
in probing. Considering the optical damping $\gamma _{fb}\approx 20Hz$ in
cooling, we also depict the sensitivity-pressure relation for the
traditional cooling scheme with the sample time of 1s by the blue dash line.
In contrast to this, our sensing displays notable advantage in precision
under low pressure environment. The sensitivity can be improved about 3
orders of magnitude than "measuring during cooling" as shown in Fig.2, and
6-7 orders than the traditional electrical method. Moveover, the resonance
frequency can be detected at a short time($\sim 26ms$) defined by our
detection bandwidth ($\sim 6Hz$). A long averaging time would lower the
signal to noise ratio. A short sample time is required to extract the signal
from the noise since the amount of energy in the burst is to be compared
with the thermal fluctuations in the same time interval[29].

\section{Conclusion}

In conclusion, we propose an ultrasensitive mass sensor using optically
trapped microdisk via the method of the "measuring after cooling". We have
demonstrated that the measurement can be finished at a short time before the
mechanical oscillator are heating out of the ground state without cooling.
In the absence of cooling damping, the sensitivity can be improved
remarkably. The resolution of $10^{-6}$Da allows us to detect precisely the
additional mass in the adsorption of an atom or molecule. Compared with
traditional electrical mass detection there are a lot of incomparable
advantages for this optical mass sensing. For example, the heat effect and
energy loss caused by circuits can be removed during the optical mass
measurement, the spectral width in optical sensing is narrower($10^{-7}$Hz)
than in electrical technique and the ultrashort measuring time($26ms$)
allows ultrahigh time resolution. The all-optical mass sensing provides a
new platform in nanoscale measurement, and has a promising to enhance the
mass sensitivity and accuracy in the future.

\begin{acknowledgments}
This work was supported by the National Natural Science Foundation of China
(Nos.11274230 and 11574206), the Basic Research Program of the Committee of
Science and Technology of Shanghai (No.14JC1491700).
\end{acknowledgments}

\end{document}